\def\kms{\hbox{km s$^{-1}$\,}}

\def\hii{H{\sc ii}\,}

\def\msun{M$_{\odot}$}
\def\mjyb{\hbox{mJy beam$^{-1}$}}

\def\cm2{cm$^{-2}$}
\def\cm3{cm$^{-3}$}

\def\ha{H$\alpha$\  }

\def\gra{$^{\circ}$}
\def\arcmin{$'$}

\def\micr{$\mu$m} 
\def\coa{\hbox{CO J=1$\rightarrow$0}}
\def\cob{\hbox{$^{13}$CO J=1$\rightarrow$0}}
\def\radec{\hbox{(R.A., Dec.)$_{ J2000}$}}
\def\radecb{\hbox{(R.A., Dec)$_{ J2000}$}}

\documentclass[printer]{aa} \usepackage{txfonts}  
\usepackage{graphicx}
\usepackage{txfonts}
\usepackage{natbib}

\begin{document}
%
 %  \title{$^{12}$CO  and $^{13}$CO J=1$\rightarrow$0 observations in the environs of the ring nebula around  WR 16}

\title{Carbon monoxide  in the environs of the star  WR 16}

\author{N. U. Duronea\inst{1,3}
          \and
           E. M. Arnal\inst{1,2}
          \and L.  Bronfman\inst{3}  
          }
\institute{Instituto Argentino de Radioastronom\'ia, CONICET, CCT-La Plata, 
 C.C.5., 1894, Villa Elisa, Argentina   \email{duronea@iar.unlp.edu.ar}\and Facultad de Ciencias Astron\'omicas y Geof\'isicas, Universidad Nacional de La Plata, Paseo del Bosque s/n, 1900 La Plata,  Argentina\and Departamento de Astronom\'ia, Universidad de Chile, Casilla 36-D, Santiago, Chile}

%  \institute{Instituto Argentino de Radioastronom\'{\i}a (IAR),
%              C. C. No 5, 1894 Villa Elisa, Argentina
%         \and
%             Facultad de Ciencias Astron\'omicas y Geof\'{\i}sicas Universidad de La Plata,
%             Paseo del Bosque s/n 1900 La Plata, Argentina}

   \date{Received ---------, 2012; accepted ------, 2012}

% \abstract{}{}{}{}{} 
% 5 {} token are mandatory
 
  \abstract
  % context heading (optional)
  % {}  
   {}
  % aims heading (mandatory)
   { We analyze the carbon monoxide emission around the star WR 16     aiming  to study  the physical characteristics of the molecular gas linked to the star   and to achieve a better understanding of the interaction between massive stars with their surroundings.    }
  % methods heading (mandatory)
   {We   study   the molecular gas in  a region  $\sim$ \hbox{86$'$.4 $\times$ 86$'$.4}   in size using \coa\ and  \cob\ line data obtained with  the 4-m NANTEN telescope.  Radio continuum archival  data at 4.85 GHz, obtained from the  Parkes-MIT-NRAO Southern Radio Survey, are also analyzed to account for the ionized gas. Available IRAS (HIRES) 60 $\mu$m and 100 $\mu$m images are  used to study the characteristics of the dust around the star.    }
  % results heading (mandatory)
   { Our new  CO and $^{13}$CO data  allow   the low/intermediate density molecular gas surrounding the WR  nebula to be completely mapped. We report two molecular features at $-5$  \kms and $-$8.5 \kms (component 1 and component 2, respectively) having a good morphological resemblance with the \ha\ emission of the ring nebula.  Component 2 seems to be associated with the external ring, whilst component 1 is placed at the interface between component 2 and the H$\alpha$ emission.  We also report a third molecular feature  $\sim$ 10$'$ in size  (component 3)  at a velocity of $\sim$9.5 \kms  having a good morphological correspondence with the inner optical and IR emission, although high resolution observations are recommended to confirm its existence.    The stratified morphology and kinematics of the molecular gas   could be associated to shock fronts and high mass-loss events related to different evolutive phases of the WR star, which have acted upon the surrounding circumstellar molecular gas.  An analysis of the mass of component 1 suggests  that this feature  is composed by swept-up interstellar gas and  is probably enriched by molecular ejecta.   The direction of the proper motion of WR 16 suggests that the  morphology observed at  infrared,  optical, radio continuum, and probably molecular emission of the inner ring nebula  is induced by the stellar motion.  }
  % conclusions heading (optional), leave it empty if necessary
   {}

   \maketitle

%________________________________________________________________

\section{Introduction}

  The interstellar medium (ISM) surrounding massive stars is expected to be strongly modified and  disturbed by the  stellar winds and ionizing radiation. There is now consensus that massive stars (\hbox{$\ge$  30 M$_\odot $})    evolve  trough a sequence of three stages, each of them  characterized by a different kind of wind \citep{gs95a,gs95b}. During the long-lived O phase ($t$ $\sim$  \hbox{(2 - 4) $\times$ 10$^6$} yr), the gas around the star is first  ionized by the high Lyman continuum flux, producing an \hii region that is later evacuated by the powerful stellar winds creating an ``interstellar bubble'' (IB) \citep{c75,D77,W77}. The latter  have been successfully detected both as thermal radio continuum shells and in the 21 cm line of atomic hydrogen \citep{ar92,cnhk96,acrc99} in the form of shells  having diameters of tens of parsecs, and expansion velocities of  \hbox{6 - 20 \kms}. In the short-lived  red supergiant (RSG) ($t$ $\sim$ \hbox{(2 - 3) $\times$ 10$^5$} yr)  and luminous blue variable (LBV) ($t$ $\sim$ \hbox{1 $\times$ 10$^4$} yr)   stages, the stellar wind becomes denser and slower and ejections of up to tens of  solar masses take place. The Wolf-Rayet (WR) stage ($t$ $\sim$ \hbox{(2 - 6) $\times$ 10$^5$} yr)    is the last  evolutionary phase  of  the massive stars prior to its explosion as a supernova, and is  characterized by  a fast and chemically enriched stellar wind which rapidly reaches and interacts with the previous RSG/LBV wind creating a ring nebula  \citep{chu81,chu81b,chu81c,chu82}.  Hence, the mutual interactions between stellar winds and ejecta of different stages may form multiple shells in the ISM at scales from tenths to tens of parsecs during the star evolution   \citep{gs96a,gs96b}.    

 Since studies of the ISM around WR stars can provide important information about the interaction  between massive stars and their environs, molecular line observations are crucial instruments to understand that interaction. The  molecular gas around WR stars  has been analyzed  in the last few years.  One of the best studied cases is the ring nebula NGC 2359 around the star WR 7  \citep{rmh01a,rmm01b,crg01} which shows clear signs  of interaction between the star and its molecular and ionized environment.  The stratification of the kinematics, morphology, and density of the molecular gas with    respect to the central star can be associated to several energetic events related with different evolutionary stages previous to the actual WR phase like RSG or  LVB phases  \citep{r203}.

\begin{figure}
\centering
\includegraphics[width=250pt]{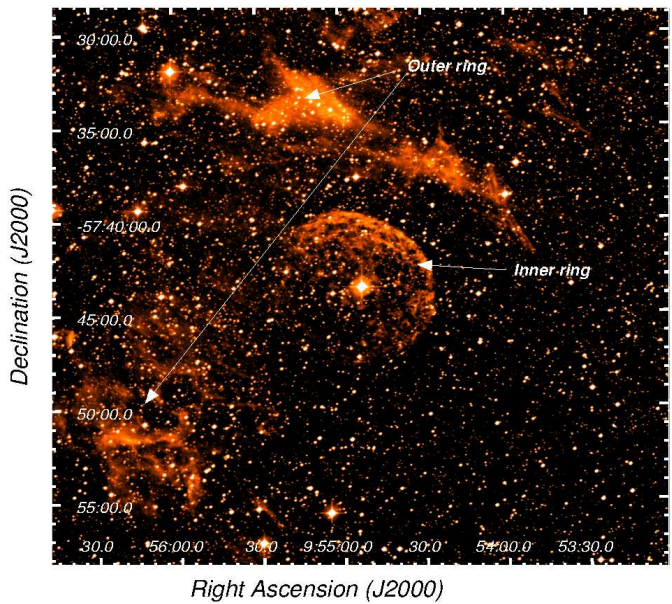}
\caption{  Super Cosmos \ha image  of the ring nebula around WR 16}.
\label{fig:wr16halfa}
\end{figure}

The \ha\ ring nebula around the WN8h star WR 16  ($\equiv$HD 86161) \citep{vdh01}, was first noticed by \citet{mcg94}. The nebula consists of a double ring system with a very intense inner ring $\sim$ 5$'$ north  from the star, whilst the  outer  ring is placed  to the northwest at $\sim$8$'$. The emission of the outer ring appears to extend to the southeast of the star. For the sake of clarity, in Fig.{\ref{fig:wr16halfa}} the \ha\ emission distribution around WR 16 obtained from the Super Cosmos \ha\ Survey (SHS)\footnote{http://www-wfau.roe.ac.uk/sss/halpha/index.html} is shown.  The stratified distribution of the \ha emission around WR 16 may be indicative of an evolution of the ISM associated with different stages prior to the WR phase, which  makes this nebula an excellent laboratory to study the interaction between  WR stars with their environs.   \citet{mwb99} surveyed the \coa\ line in a grid of  $\sim$20$'$$\times$14$'$ centered at WR16 using SEST (HPBW = 45$''$ ). They found a molecular ``cocoon'' at a radial velocity\footnote{Radial velocities are referred to the local standard of rest ({\rm LSR})} of $\sim$ $-$5 \kms around WR 16. They estimated a total  molecular mass in the range \hbox{5 - 78 M$_{\odot}$}. This fact, together with a chemical enrichment observed   towards the inner part of the nebula, led the authors to suggest an stellar origin for both, the inner nebula and the molecular gas. 

 Spectrophotometric distances of WR 16 were estimated by several authors: 2.6 kpc \citep{cv90}, 2 kpc \citep{mwb99}, 2.37 kpc \citep{vdh01}. In this paper we adopt an averaged value of  2.3 $\pm$ 0.3 kpc.

Although providing important information about the molecular gas associated with  the close surroundings around WR 16, the observations of \citet{mwb99}   do not provide an  entire picture of the    molecular environment of the nebula because they observed the distribution of the  molecular gas in a  limited  region around the WR star.  Clearly, the CO emission distribution extends well beyond the area surveyed by these authors.       In order to complete the analysis of \citet{mwb99}, we study the molecular gas over a region $\sim$ 86$'$ $\times$ 86$'$ around the star  using observations of the J=1$\rightarrow$0 transition of the   CO   and $^{13}$CO molecules. Our aim is  to investigate in more detail the morphology, kinematics and physical properties of the  molecular environment of WR 16. To account for the properties of the ionized gas and dust we analyze available archival radio continuum and infrared (IR) data.      The observations and databases used in this work are outlined in Sect. 2, the results are described in Sect. 3, and the discussion in Sect. 4. Concluding remarks are  presented in \hbox{Sect. 5.}

%\begin{figure*}
%\centering
%\includegraphics[width=430pt]{wr16perfiles2.eps}
%\caption{Mean CO emission profiles toward six regions around WR 16. The CO profiles are averaged over a square area $\sim$ 3$'$ in size, centered on the black dots drawn on the H${\alpha}$ image (center). The profile units are T$^{*}_R$ (ordinate) and V$_{LSR}$ en \kms.    }
%\label{fig:wr16ispec}
%\end{figure*}

%__________________________________________________________________

\section{Observations and data bases}

\subsection{CO data}
  
 The \coa\ data  were obtained using the 4-m {\rm NANTEN} millimeter-wave telescope of Nagoya University. At the time  we  carried out the observations (April 2001), this telescope was installed at Las Campanas Observatory, Chile. The half-power beamwidth and the system temperature, including the atmospheric contribution towards the zenith, were  2$'$.6 ($\sim$1.7 pc at 2.3 kpc) and $\sim$ 220\,K (SSB) at 115 GHz, respectively. The data were gathered using the position switching  mode with absolute reference positions; the points devoid of CO emission, used as reference, were available at the observatory and observed interspersed among the program positions. The spectrometer used was  acousto-optical with 2048 channels  with a   frequency resolution of 20 kHz (equivalent to 0.055 \kms at 115 GHz) which  provides a total coverage of $\sim$113 \kms. For intensity calibrations, a room-temperature chopper wheel was employed \citep{pb73}. An absolute  intensity calibration  \citep{uh76,ku81} was achieved by observing Orion {\rm KL}  and $\rho$ Oph East.   The absolute radiation temperatures, $T_{\rm R}^\ast$, of Orion {\rm KL} and $\rho$ Oph East, which were both observed by the {\rm NANTEN} radio telescope, were assumed to be 65 K and 15 K, respectively \citep{myom01}.   A second-order degree polynomial was subtracted from the observations to account for instrumental baseline effects. The spectra were reduced using CLASS software\footnote{ htto://www.iram.fr/IRAMFR/PDB/class/class.html}.  The {\rm CO} observations were centered at ({\it l,b}) =  (281$^{\circ}$04$'$47$''$ -02$^{\circ}$33$'$03$''$) (\radecb\ = \hbox{09$^h$54$^m$53$^s$ -57\gra43\arcmin38\arcsec})        with two areas covered with different samplings.  An inner area of  \hbox{$\Delta$$l$ $\times$ $\Delta$$b$} =  \hbox{32$'$.4 $\times$ 32$'$.4}  was sampled one beam apart, whilst a larger area of \hbox{$\Delta$$l$ $\times$ $\Delta$$b$} = \hbox{86$'$.4 $\times$ 86$'$.4} was sampled every 5.$'$4 (two beamwidth). This technique allowed us to obtain two datacubes covering different areas around WR 16. The integration time per point was 16s resulting in a typical rms noise of $\sim$ 0.3 K.

 \subsection{$^{13}$CO data} 

   The radio astronomical group at Nagoya University has carried out a \cob\  survey of the Galactic Plane using the NANTEN telescope. In this work we use part of this unpublished survey that was kindly provided to us by the  group. The beam size of the telescope at $^{13}$CO frequency (110 GHz)   is 2.$'$7   and  the typical rms noise is $\sim$0.2 K. The frequency resolution is 40 kHz (equivalent to 0.1 \kms at 110 GHz).   Observations were made in position switching mode and at a grid spacing of 4$'$

\subsection{Complementary data}

The CO and $^{13}$CO data are complemented with narrow-band H${\alpha}$ data retrieved from the SuperCOSMOS H-alpha Survey (SHS). The images have a sensitivity of 5 Rayleigh (1 Rayleigh = 2.41 $\times$ 10$^{-7}$ erg cm$^{-2}$ s$^{-1}$ sr$^{-1}$),  and $\sim$1$''$ spatial resolution \citep{p05}. We  also include radio continuum 4.85 GHz data  retrieved from the Parkes-MIT-NRAO (PMN) Southern Radio Survey.   The images have $\sim$5$'$ resolution and $\sim$8 \mjyb\  total rms noise   \citep{c93}. To study the characteristics of the dust in the nebula we use infrared images  at 60 \micr\  and 100 \micr\   retrieved from the  high resolution IRAS images (HIRES) database\footnote{http://irsa.ipac.caltech.edu/applications/IRAS/IGA/} \citep{fa94}.

%\radec\ = \hbox{(9$^h$55$^m$45$^s$,--58\gra 40\arcmin)}. 

\section{Results and analysis of the observations}

\subsection{Distribution and physical properties of the CO   and  $^{13}$CO}

We investigate the presence of molecular gas in the environs of WR 16 using several  CO spectra obtained towards different position around the star.  After a careful inspection to the individual spectra, we note that the bulk of the CO emission is concentrated in the velocity range from $\sim$ $-$10 \kms to $\sim$ $-$3 \kms. We use a mean radial velocity of $\sim$ $-$6   \kms to make a rough estimate  of the kinematical distance. Using the analytical fit to the circular galactic rotation model  of \citet{bb93} along \hbox{{\it l} $\approx$ 281\gra} we estimate near and far  kinematical distances of about  0.6  kpc   and  2.8 kpc, respectively.  Hence, the far kinematical distance is in agreement (within errors) with the adopted distance  for WR 16. This is indicative of a likely association between the molecular gas detected in the mentioned velocity range  and the WR star.  We have to keep in mind, however, that non-circular motions of up to 8 \kms are know to exist in the Galaxy \citep{bg78,bb93} which are of the same order of the velocity of the molecular feature. Although this  might introduce large uncertainties in the kinematical distance determination it suffices for discerning between near and far kinematical distance.

 We have analyzed the CO and $^{13}$CO data cubes in velocity intervals of 0.2 \kms whithin the mentioned velocity range. This  strategy allowed us to identify three molecular features that are morphologically correlated with different regions of the ring nebula. The central velocities of these features are approximately at   $-$5,  $-$8.5, and  $-$9.5 \kms.  With the aim to facilitate further analysis we shall refer to these features  as component 1,  component 2,  and component 3, respectively. In the following, we describe and analyse the three features in detail.

\subsubsection{Component 1}

\begin{figure}[h!]
\centering
\includegraphics[width=210pt]{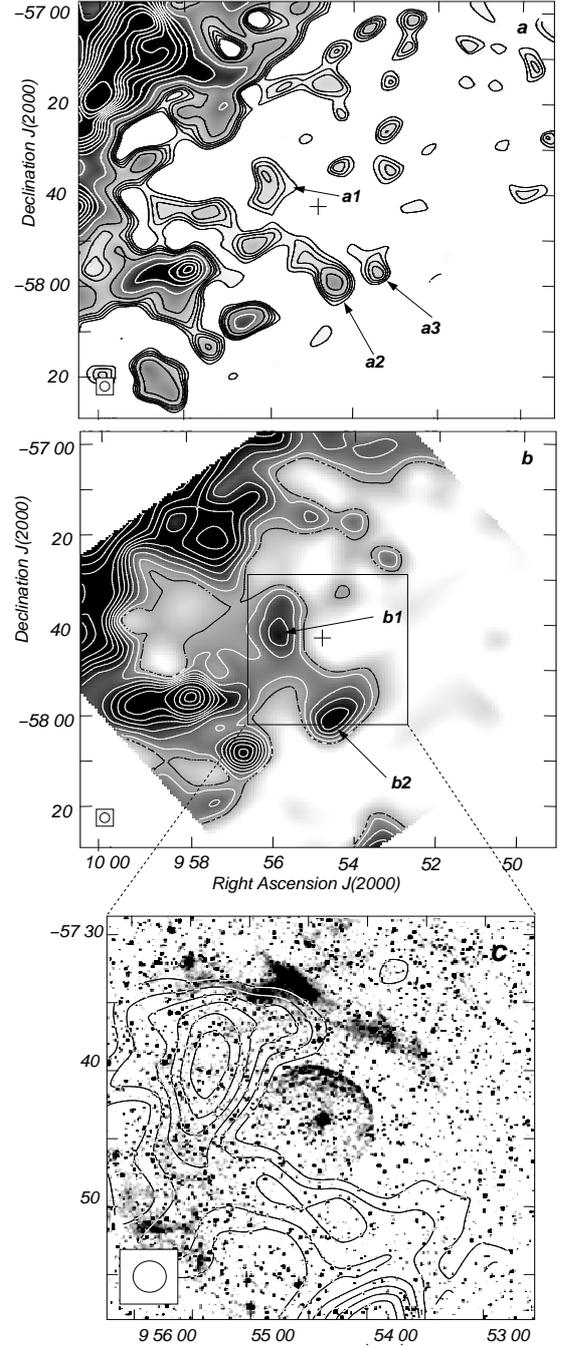}
\caption{ Distribution of the mean CO emission integrated in the velocity range from $-$7.2 \kms to $-$3.2 \kms. The beam size of the CO observations is shown by a white circle  in the lower left corner of each image. The position of WR 16 is marked by a plus sign at the center of each image.  a) Emission distribution of the \cob.  The lowest temperature contour is 0.16 K ($\sim$ 5 rms). The contour spacing temperature is 0.08 K till 0.48 K, and here\ onwards\ the contour\ spacing is  0.35 K.   b) Emission distribution of the \coa.  The lowest temperature contour is 0.8  K ($\sim$10 rms) and the contour spacing temperature is 0.7 K.  c)  Overlay of the mean CO  emission  in the  velocity range from $-$7.2 \kms to $-$3.2 \kms    (contours)  and the  \ha\ emission (grey scale) in  the central region. Contour levels are the same as in Fig 2b.       }
\label{fig:12y13}
\end{figure}

In Figs. \ref{fig:12y13}a and \ref{fig:12y13}b    we show the emission distribution of the $^{13}$CO and CO lines, respectively, in the velocity range from   $-$7.2 \kms to $-$3.2 \kms. Although the  emission lines of CO and $^{13}$CO   trace the molecular gas of different densities, a molecular feature   close to WR 16 stands out in both figures. This feature engulfs a low molecular emission region centered at the position of WR 16,  surrounding the star from east to south. Fig. \ref{fig:12y13}b also shows that  the CO feature  appears to be projected onto a more extended CO plateau  having   a relatively strong ($\sim$10 rms) emission, that extends towards a large and bright  molecular  structure  lying $\sim$20'  to the north and east, which is likely  part of the galactic disk. %The plateau is not observed in the $^{13}$CO emission (Fig. \ref{fig:12y13}a), with the exception of  two small clouds  placed  at  \radec\ = \hbox{(9$^h$57$^m$30$^s$,--57\gra 42\arcmin)}    and \radec\ = \hbox{(9$^h$56$^m$30$^s$,--57\gra 51\arcmin)}.  

In order to follow in more detail the morphology  of component 1  and to compare with the  H$\alpha$ emission, we show  in Fig. \ref{fig:12y13}c an overlay of the CO emission  in the velocity range from  $-$7.2 \kms to $-$3.2 \kms obtained with the 32$'$.4 $\times$ 32$'$.4 datacube, with the SHS emission of the ring nebula.  Differences in the contour lines between Figs. \ref{fig:12y13}b  and \ref{fig:12y13}c are due to the different  sampling of both  datacubes.  This molecular feature was  already reported by \citet{mwb99}, referred to in that work as ``molecular cocoon''.

We identify a number of clouds  in the CO and $^{13}$CO emissions. They are labeled as {\it a1},  {\it a2}, and  {\it a3} for the $^{13}$CO map, whilst for the CO image they are labeled as {\it b1} and {\it b2}. This identification is arbitrary and was made only to facilitate further analysis. Cloud {\it a1} is very likely the $^{13}$CO counterpart of  {\it b1}. On the other hand  {\it a2}, and  {\it a3} are possibly the counterparts of  {\it b2} which  appear as a single structure due to the worse CO sampling.  Assuming that  the \cob\ line is in local thermodynamic equilibrium (LTE),   we follow the method described by \citet{d78} to  calculate some physical parameters such as the peak optical depth, $\tau(^{13}$CO) and the  molecular column density, $N(^{13} \rm CO)$. To use this method we assume that $\tau(^{13}$CO) $<<$ 1, $\tau$(CO) $>>$ 1, and $T_{\rm exc}$(CO) = $T_{\rm exc}$($^{13}$CO).       Afterwards, we calculate the total  hydrogen  mass, $M(\rm H_2)$, via
\begin{equation}\label{eq:masa}
  M(\rm H_2)\ =\   (m_{sun})^{-1}\  \mu\ m_H\ \sum\ \Omega\ {\it N}(\rm H_2)\ {\it d}^2 \quad  \quad \quad   \textrm{(M$_{\odot}$)}
\end{equation}
where  m$_{\rm sun}$ is the solar mass ($\sim$ 2 $\times$ 10$^{33}$ g),    $\mu$ is the mean molecular weight, which is  assumed to be equal to 2.8 after allowance of a relative helium abundance of 25\% by mass \citep{Y99},  m$_{\rm H}$ is the hydrogen atom mass   ($\sim$ 1.67 $\times$ 10$^{-24}$ g), $\Omega$ is the solid angle subtended by the CO feature  in ster, $d$ is the assumed  distance expressed in cm, and  $N$(H$_2$) is the H$_2$ column density, obtained  using a ``canonical'' abundance \hbox{$N(\rm H_2)$/$N(^{13} {\rm CO})$} = \hbox{5 $\times$ 10$^{5}$} \citep{d78}. We also estimate the optical depth of the \coa\ line from the \cob\ line with  $\tau(\rm CO)$$\approx$$(\rm CO/ ^{13}CO)$$\tau(^{13} \rm CO)$, 
%\begin{eqnarray}
%\tau(\rm CO)\ &  =\ &  [\nu(^{13} \rm CO)/\nu(\rm CO)]^2  \\
% && \times\:  [\Delta v(^{13}\rm CO)/ \Delta  v(\rm CO)]\ (\rm CO/ ^{13}CO)\    \tau(^{13} \rm CO) \nonumber
% \end{eqnarray}
%\begin{equation}
%\quad \tau(\rm CO)\ =\  \left[\frac{\nu(^{13} \rm CO)}{\nu(\rm CO)}  \right]^2\  \left[\frac{\Delta v(^{13}\rm CO)} {\Delta  v(\rm CO)} \right]\ \left(\frac{\rm CO}{^{13}CO} \right)\   \tau(^{13} \rm CO)
%\label{eq:tau12}
%\end{equation}
where  CO/$^{13}$CO is the isotope ratio (assumed to be $\sim$62; \citealt{lp93}). It is worth to point out that we adopted peak values of $T^*_R(^{13}\rm CO)$, which yields to upper limits for the line opacities, column densities, and  molecular masses.

The mass of the molecular gas can also be estimated using the \coa\ line via two  methods: {\it (a)} Using the relation between  integrated emission of the \coa\ line, \hbox{$I$(CO)} = $\int\ T^*_{\rm R} {\rm \tiny{(CO)}}   \ d{\rm v}$, and molecular hydrogen column density,  $N({\rm H_2})$,  and {\it (b)}    the virial theorem.     For the first method we use
\begin{equation}\label{eq:X}
 \quad   N({\rm H_2})\ =    {\rm X} \  \times\  I{\rm (CO)} \ \ \ \qquad  \qquad  \qquad \qquad ({\rm  cm}^{-2})
\end{equation}
The X-factor  lies in the range \hbox{(1 - 3)} $\times$ 10$^{20}$ cm$^{-2}$ (K \kms)$^{-1}$, as estimated by the virial theorem and $\gamma$-ray emission   \citep{blo86,sol87,ber93,d96,sm96}. In this paper we adopt X = 1.6 $\times$ 10$^{20}$ cm$^{-2}$ (K \kms)$^{-1}$ \citep{hu97}. For the second method, considering only gravitational and internal pressure  and  assuming a spherically symmetric cloud with a $r^{-1}$ density distribution, the molecular mass can be determined from
\begin{equation}\label{eq:virial}
\quad M\ =\ 190\ R\ (\Delta  v_{\rm cld})^2  \qquad \qquad   \textrm{(M$_{\odot}$)}
\end{equation}
\citep{ml88}, where $R$=$\sqrt{A_{\rm cloud}/ \pi}$ is the effective radius in parsecs,  and $\Delta v_{\rm cld}$ is   defined as the FWHM line width of the composite profile derived by using a single Gaussian fitting. The composite profile is obtained by averaging all the spectra within the area of the cloud ($A_{\rm cloud}$).  The uncertainties involved are about \hbox{40 $\%$}, mostly due to the estimates of {\it R}.  In Table \ref{tabla:1} we summarize some of the physical parameters detailed above, obtained for the molecular clouds  constituting  component 1. The values of $N$(H$_2$)$_{\rm X}$ and $M({\rm H_2})_{\rm X}$ are the hydrogen column density and molecular mass, respectively, determined using Eq.\ref{eq:X}, whilst $N(^{13} \rm CO)_{\rm LTE}$ and  $M(\rm H_2)_{LTE}$ are the $^{13}$CO column density and molecular mass obtained assuming LTE. The virialized mass is given by  $M_{\rm vir}$.  In Table \ref{tabla:1} we also  list the thermal line widths for {\it b1} and {\it b2}, $\Delta v_{\rm th}$, obtained assuming $T_{\rm kin}$ = $T_{\rm exc}$ (LTE).       To derive the total molecular mass of component 1  from the  $^{13}$CO data we added up the values   obtained for {\it a1}, {\it a2}, and {\it a3}, which yields to a total mass   $M(\rm H_2)_{LTE}$ $\approx$ 900 \msun. Using the same technique with the values obtained for  {\it b1} and {\it b2}  from the CO data, we derive   total molecular masses  $M({\rm H_2})_{\rm X}$ $\approx$ 5500 \msun\ and   $M_{\rm vir}$ $\approx$ 16000 \msun. 

 It is worth to point out that the physical parameters quoted in Table \ref{tabla:1} are only derived for  component 1, since is the only one entirely detected at $^{13}$CO emission and  morphologically defined without ambiguity.

\subsubsection{Component 2}

\begin{figure}[h!]
\centering
\includegraphics[width=220pt]{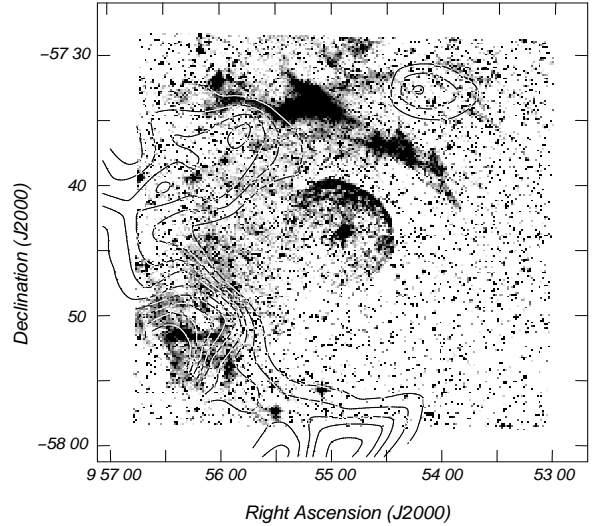}
\caption{  CO integrated intensity  in the  velocity   range from  $-$9.5 \kms  to $-$7.2 \kms (contours) superimposed onto the \ha emission (grey scale). The contour levels start at 0.8 K and the contour spacing is 0.55 K.   }
\label{fig:comp2halfa}
\end{figure}

In Fig. \ref{fig:comp2halfa} we show the spatial distribution of the CO emission in the velocity interval  from $-$9.5 \kms to $-$7.2 \kms, overlaid on the \ha\ emission of the nebula. The molecular gas in this velocity interval  shows a  morphological correspondence with the outer \ha\ ring, especially at the intense optical emission at \radec\ = \hbox{(9$^h$56$^m$15$^s$,--57\gra 51\arcmin)}.  This feature was not detected by \citet{mwb99} due to the limited area covered by their CO data.  It can be noted that this feature extends well beyond the area depicted in Fig. \ref{fig:comp2halfa} which suggests that is part of a larger structure, probably the previously reported extended CO plateau lying aside component 1   (see  Fig. \ref{fig:12y13}b). 

\begin{figure*}%[h!]
\centering
\includegraphics[width=470pt]{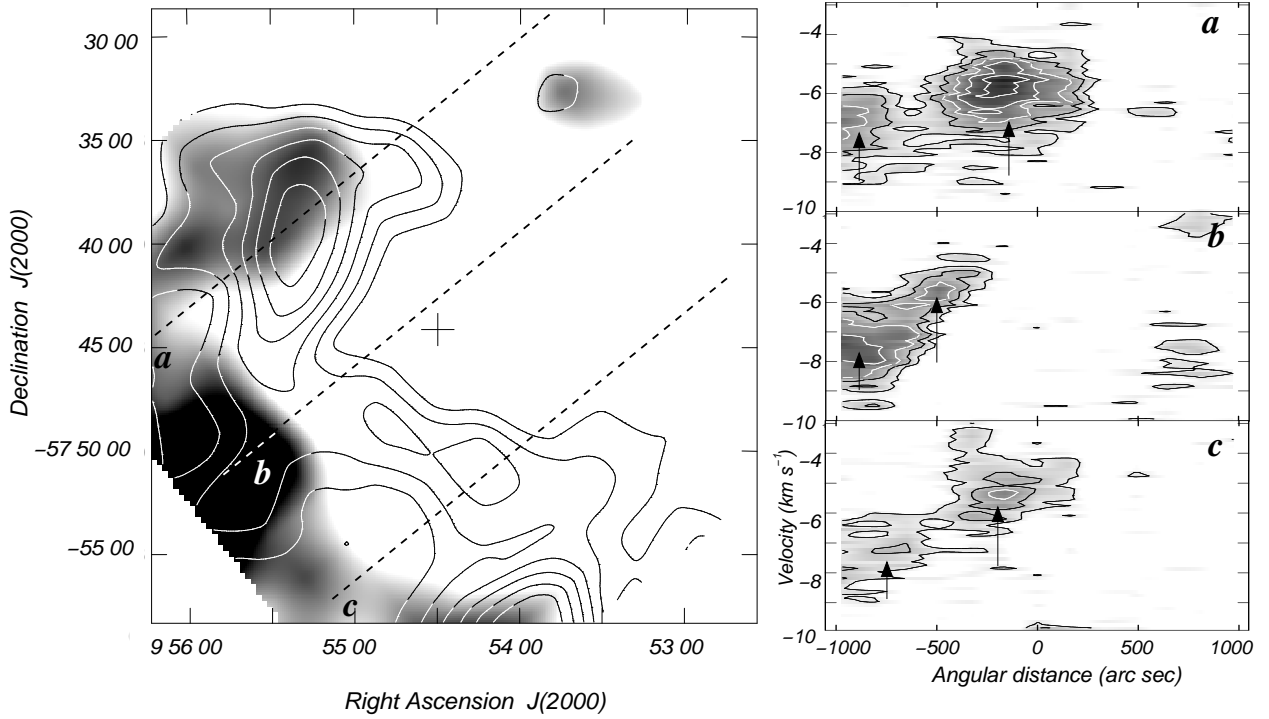}
\caption{ {\it Left panel:} Comparison between  component 1 (contours)   and component 2 (grey scale). The location of the slices, labeled as  {\it a}, {\it b}, and {\it c}, are sketched by the dotted lines. The  cross at the center indicates the position of WR 16. {\it Right panels:} Position-velocity maps obtained for strips  {\it a}, {\it b}, and {\it c}. Contour levels start at 1.9 K and the contour spacing is 1.4 K. The loci of both components are indicated by the black arrows.  }
\label{fig:comp1y2}
\end{figure*}
In Fig. \ref{fig:comp1y2} we show the  the CO emission distribution  of component 2    (grayscale)       superimposed on that of component 1     (contours).    A spatial separation between both components can be observed, with component 2 adjacent to the eastern  border of the outer H$\alpha$ ring, and component 1 placed almost at the interface between the latter and the H$\alpha$ emission. The kinematic of the molecular gas was analyzed by position-velocity maps across selected strips. In the right panels of Fig. \ref{fig:comp1y2} we show the position-velocity diagrams in the directions sketched by the dotted lines in the left panel of the figure. Slices {\it a} to {\it c}  show that the  peak emission of both features are discernible  and spatially  separated by about $\sim$ 7$'$. Both slices also show the presence of a weak bridge of CO emission which connects the western border of component 1   with the eastern side of component 2. Although this bridge might suggest a physical link between both features, no hints of a velocity gradient is detected. On the other hand, slice {\it c} shows a stronger CO emission bridge between the two features. However, the peak emission of both features are well separated by $\sim$ 6$'$. Probably, the selected direction of slice {\it b}, along the stronger region  of component 2, and the weaker region of component 1, makes their emissions  to look merged,   although this may be due  to a projection effect. The morphology and kinematical  behavior of component 2    suggests that the presence  of this feature   is not consequence of a mere velocity gradient, but to the presence of a second molecular structure. As  suggested  before, component 2 may be  a small part of a larger molecular structure (plateau), whose extension likely exceeds the area depicted in Figs. \ref{fig:comp2halfa}   and  \ref{fig:comp1y2}. The double-layer morphology reported above for the molecular gas in the velocity range from $\sim$ --10 \kms to $\sim$ --3 \kms could be associated to some shock fronts and  mass-loss rate episodes  which have been acting  upon the surrounding ISM, as  predicted by the hydrodynamical models of \citet{gs96a,gs96b}. Characteristics times of this features are further discussed in Sect. 4.2.

\subsubsection{Component 3}

\begin{figure}[h!]
\centering
\includegraphics[width=220pt]{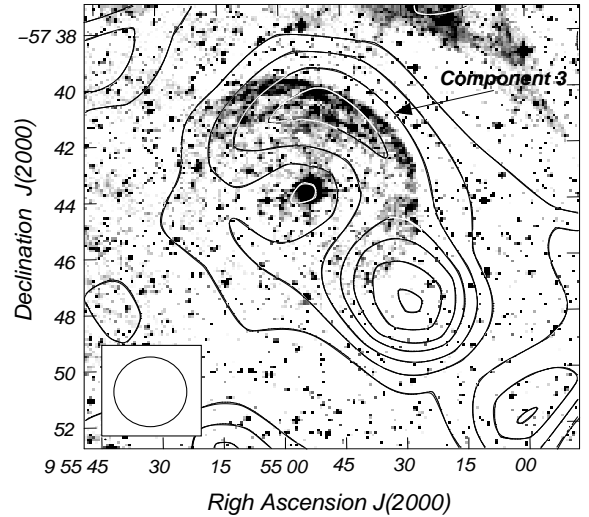}
\caption{  Distribution of the mean CO emission in the velocity interval from --9.59 \kms to --9.48 \kms (contours),  superimposed on the \ha\ emission (grey scale).  The beam size of the CO observations is shown by a white circle  in the lower left corner. The lowest temperature contour is 0.5  K ($\sim$ 4.5 rms). The contour spacing is $\sim$0.65 K.  }
\label{fig:comp3-halfa}
\end{figure}

 In Fig. \ref{fig:comp3-halfa}     we show  an   image $\sim$ 16$'$ $\times$ 16$'$ in size centered at the position of WR16 that shows  the CO emission in the velocity range from -9.59 \kms   to -9.48 \kms (two channel maps), superimposed onto the \ha\ emission of the inner ring nebula.  
 
Even though the  intense CO clump located at    \radec\ = \hbox{(9$^h$55$^m$30$^s$, --57\gra 47\arcmin 30\arcsec)}  stands out, a fainter ring-shaped molecular emission  (indicated in the figure as component 3)   surrounding the position of WR 16 is noticeable. The star is seen projected onto a well-defined minimum of the CO emission whilst the  molecular gas  $\sim$ 3$'$ north of the star is coincident with the brightest \ha\ emission  of the inner ring.  
The good morphological correspondence  between the CO and the \ha emission may be  indicative of a mutual physical association between the molecular and the ionized gas.  It is worth to mention that this feature was not detected by \citet{mwb99},  probably due to sensitivity or  velocity resolution. However, we are cautious about reporting component 3  since the area is small and the  CO observations are not fully-sampled. Furthermore, we note the short velocity interval in which component 3 is noticeable. Further high-spatial resolution observations with instruments like APEX may help to clarify the nature and characteristics of this feature.

\begin{table}%[h!]
\caption{ Main physical parameters of the identified molecular clouds}
\label{tabla:1}
\begin{center}
\begin{tabular}{l c c c  }
\hline
 &     $^{13}$CO clouds   & &    \\
\hline\hline              
Parameters      &  {\it a1}  &  {\it a2}   &  {\it a3}      \\
\hline
%&&&\\
Center position: &&&\\
RA$_{\rm (J2000)}$ ($^{h}$ $^{m}$ $^{sec}$)  &   09 56 30 	    &  09 54 30      &   09 53 00      \\

Dec$_{\rm (J2000)}$ (\gra\ \arcmin\ \arcsec)  &   -57 37 30 	  &    -57 59 30       &   -57 57 00     \\
 $T^*$$_{\rm R}$(Peak) (K)  &  0.45  &  0.53   &  0.32   \\
% T$^*$$_{\rm R}$(average)  (K)  &  0.17    &  0.22   & 0.06   \\ 
$\tau$($^{13}$CO)   &  $\sim$0.09  &  $\sim$0.11   &  $\sim$0.07    \\

$\tau$(CO)  &  $\sim$1.7  &  $\sim$2.5   &  $\sim$1.1    \\
$\Omega$ (10$^{-6}$ ster)   &  7.4  & 9.7 & 5.1 \\ 
%$N$($^{13}$CO)$_{\rm LTE}$(10$^{15}$ cm$^{-2}$)  &  1.5  & 2.8   &  1.0    \\
$N$($^{13}$CO)$_{\rm LTE}$(10$^{15}$ cm$^{-2}$)  & $\lesssim$   0.7  & $\lesssim$  1.1   & $\lesssim$   0.15    \\
$N$(H$_2$)$_{\rm LTE}$ (10$^{20}$ cm$^{-2}$) &  $\lesssim$ 3.4  &  $\lesssim$  5.3   &  $\lesssim$  0.8    \\
$M({\rm H_2})$$_{\rm LTE}$ (10$^2$ M$\odot$)  &   $\sim$ 2.8  &  $\sim$  5.6     & $\sim$  0.5    \\
%$n({\rm H_2})$ (cm$^{-3}$) &  130$\pm$70  &  760$\pm$350   &  340$\pm$160    \\
%&&&\\
\hline
  &   CO clouds   & &    \\
\hline\hline              
Parameters        &  {\it b1}    & {\it b2} &  \\
\hline
%&&&\\
Center position: &&&\\
RA$_{\rm (J2000)}$ ($^{h}$ $^{m}$ $^{sec}$)   &  09 56 00  &  09 54 30   &     \\
Dec$_{\rm (J2000)}$ (\gra\ \arcmin\ \arcsec)   &  -57 43 00    &   -58 02 00    &    \\
$T^*$$_{\rm R}$(Peak)  (K) &  3.9  &  4.1   &     \\
$T_{\rm exc}$ (K)  &  7.1   &  7.3    &   \\
$T^*$$_{\rm R}$(average)    (K)  &  2.1  &  2.2   &     \\ 

$I({\rm CO})$ (K \kms)  & $\sim$ 6.3  &  $\sim$ 6.6   &   \\
$\Omega$ (10$^{-6}$ ster)   &  22.8  &  24.3  &    \\ 

$N$(H$_2$)$_{\rm X}$ (10$^{20}$ cm$^{-2}$) &  10.1   &  10.6    & \\
$M({\rm H_2})_{\rm X}$     (10$^3$ M$\odot$)   &   2.6$\pm$1.1   &    2.9$\pm$1.3     &      \\

%$N$(CO)$_{\rm LTE}$ (10$^{16}$ cm$^{-2}$)  &  3.8  & 4.1   &      \\
%$N$(CO)$_{\rm LTE}$ (10$^{16}$ cm$^{-2}$)  &  1.5  & 1.6   &      \\

%$N$(H$_2$)$_{\rm LTE}$ (10$^{20}$ cm$^{-2}$) &  1.5   &  1.6   & \\
$\Delta v_{\rm cld}$   (\kms)              &  3.3                    &  4.1             & \\
$\Delta v_{\rm th}$ (\kms) &   0.10     &   0.11        & \\
$R$ (pc)             &           2.9          &   3.1           &  \\
$M_{\rm vir}$     (10$^3$  M$\odot$)   & $\sim$6.1        & $\sim$9.9        &      \\

%&&&\\

\hline

\end{tabular}
\end{center}
\end{table}

\subsection{Radio continuum emission}

To show the distribution of the ionized gas, we display in   Fig. {\ref{fig:5ghz} the 4.85 GHz radio continuum emission image obtained from the  PMN survey, overlaid onto the SHS \ha\ emission  image of the nebula. As expected, both  images show a relatively good morphological correspondence, since  both emissions are indicative of the existence of ionized gas.  Four radio continuum features are identified within the area observed at CO frequencies. Feature 1 peaks at     \radec\ $\approx$ \hbox{(9$^h$54$^m$40$^s$,--57\gra 40\arcmin 56\arcsec)}      and has an   arc-like  morphology almost coincident with  the brightest section of the inner ring nebula. Feature 2, peaking at \radec\ $\approx$ \hbox{(9$^h$56$^m$40$^s$,--57\gra 51\arcmin 30\arcsec)}, is placed alongside a relatively strong optical structure. The peak emission of Feature 3  \radec\ $\approx$ \hbox{(9$^h$55$^m$00$^s$,--57\gra 35\arcmin 00\arcsec)}    is coincident with the brightest section of the outer ring nebula. Unlike the other radio continuum  features, Feature 4 is seen projected onto a weak optical emission. Its location and morphology suggests that this feature  might be the radio continuum counterpart of the south-western section of the outer ring nebula.

In Table \ref{tabla:2} we summarize some parameters of the ionized gas estimated from the radio continuum flux densities at 4.85 GHz ($S_{4.85}$). Adopting an electron temperature of 10$^4$ K, assuming constant electron densities, and using the spherical models of  \citet{mh67} for features 2 and 3, and their cylindrical model for features 1 and 4,    we obtained the rms electron density ($n_{\rm e}$) and the ionized mass  ($M_{\rm ion}$)  for each feature. 
\begin{figure}[h!]
\centering
\includegraphics[width=220pt]{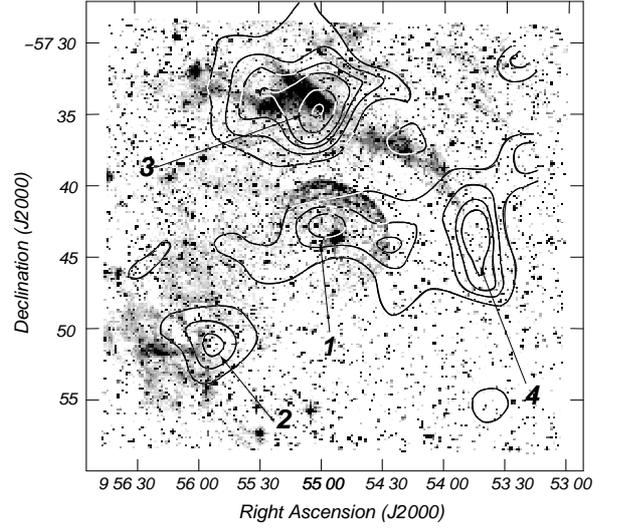}
\caption{Overlay of the radio continuum image at 4.85 GHz (contour lines) and the \ha image (grey scale). The lowest contour line is 28 \mjyb \  ($\sim$ 3.5 rms), and the  contour spacing is 50 \mjyb.   }
\label{fig:5ghz}
\end{figure}
For $S_{4.85}$ we estimate an error of about 20$\%$.  Since no clumping factor was taken into account, the obtained values of $n_{\rm e}$ and $M_{\rm ion}$ result in lower and upper limits, respectively.     To calculate  the number of Lyman  continuum photons needed to keep  the current level of ionization in the ring nebulae ($N_{\rm Lyc}$), we use the relation of \citet{ch76}. 
\begin{eqnarray}
N_{\rm Lyc} = 0.76\ \times\ 10^{47}\ T_{4}^{-0.45}\ \nu^{0.1}\ S_{\nu}\ d^2, 
\label{lyman}
\end{eqnarray}
where $T_{4}$ is the electron temperature in units of 10$^4$ K, $\nu$ is the frequency in units of GHz,   $d$ is the distance in kpc, and  $S_{\nu}$ is the total flux density in Jy. Summing up the values obtained for all the radio continuum features we obtain \hbox{$N_{\rm Lyc}$$\approx$1 $\times$ 10$^{47}$ s$^{-1}$}.  Considering that the number  of Lyman continuum photons emitted  by a WN8 star is $\sim$ 1.25 $\times$ 10$^{49}$ s$^{-1}$ \citep{cr07},  we therefore conclude that WR 16 is by far capable of maintaining the observed ionization level. 
\begin{table}[h!]
\caption{ Main parameters of the ionized gas in features 1, 2, 3, and 4}
\label{tabla:2}
\begin{center}
\begin{tabular}{ l c  c c c }
\hline
 %&\\              
features & 1  &  2 & 3 & 4 \\
%&\\
\hline\hline
%&&&&\\
$S_{4.85}$  (mJy)  &  $\sim$48 &  $\sim$24  & $\sim$95  &  $\sim$50\\
$n_{\rm e}$ \ \   (cm$^{-3}$)              &  16 &  10  & 7  & 8 \\
$M_{\rm ion}$ \ \ (\msun)             &  4 & 3   & 20  & 9 \\ 
$N_{\rm Lyc}$ \ \  (10$^{47}$ seg$^{-1}$)   &  0.2 &  0.1  & 0.5  & 0.2 \\
%&&&&\\
\hline

\end{tabular}
\end{center}
\end{table}

\section{Discussion}

\subsection{ The molecular mass of component 1 }
 
In Sect. 3.1 we have derived the mass of the molecular clouds constituting component 1 making use of the CO and $^{13}$CO data by three different methods. In spite of their uncertainties, all three  methods provide mass estimates for this feature that  strongly suggest that  mostly of  the molecular gas has a non-stellar origin. We  note that the total LTE mass obtained in this work  ($\sim$1200 \msun)  is substantially larger than that obtained by  \citet{mwb99}, although this difference may be due to the limited area covered by these authors. 

According to the obtained values a   remarkable result is noticed when comparing all three methods, namely $M_{\rm vir}$ $>$  $M({\rm H_2})_{\rm X}$ $>$  $M(\rm H_2)_{LTE}$. Regarding to the total virialized mass, the derived value ($\sim$ 16000 \msun) was obtained assuming that only the  gravitational and  internal pressure support the stability of the cloud ignoring  the effect of external pressure.  Mass estimation methods such as the Virial and CO integrated emission  masses depend on the line width, which are highly sensitive to cloud kinematics. Hence, comparing masses estimated by these methods enables some analysis of turbulence and kinematics of the molecular gas. From  Eqs. \ref{eq:X} and \ref{eq:virial} it can be noted that %$M({\rm H_2})_{\rm X}$ $\propto$ $\Delta$ v $R^2$, and $M_{\rm vir}$ $\propto$  $\Delta v^2$ $R$, respectively. Therefore, 
\hbox{$M_{\rm vir}$/$M({\rm H_2})_{\rm X}$ $\propto$ $\Delta v_{\rm cld}$/$R$}.  Therefore,  molecular  clouds having high line widths  should exhibit large virialized mass in comparison with its integrated mass. This could be the case of component 1, since {\it b1} and {\it b2} exhibit line widths significantly higher than the predicted thermal width (see Table \ref{tabla:1}). This is not  odd since the environs of a massive star is highly disturbed by  strong stellar winds and  violent   shock fronts.  The high value of $M_{\rm vir}$    with respect to $M({\rm H_2})_{\rm X}$  indicates that component 1 has too much kinetic energy and is not in virial equilibrium. Very likely, the molecular gas of component 1 is supported by  the external pressure provided by the warm and low density gas  that is ionized by the strong UV radiation field of the WR star. The pressure needed to sustain the molecular gas could be provided by the ionized gas of the ring nebula trapping \hbox{component 1}.

 The  discrepancy between  $M({\rm H_2})_{\rm X}$ and  $M(\rm H_2)_{LTE}$ is harder to explain.   The X-factor was  obtained empirically  towards  several molecular clouds in the Galactic disk, and is believed to be accurate to a factor of 2, when averaged over a large number of clouds. However, this factor can vary under conditions differing from those of typical Galactic clouds, namely  density, age, turbulence,   cosmic ionization rate, incident radiation field, and    metallicity \citep{bell06,pin08,sha11}.  The last two factors may have played a major role in the environs of  the star. In fact, these two factors are intimately related since the formation of CO is sensitive to both, the amount of available C and O, and the UV radiation field responsible for photodissociating molecules and regulating the CO abundances. Increasing metallicity could strongly decrease the X-factor   below the canonical value, whilst a stronger radiation field could slightly raise it \citep{bell06}. The chemical abundance of the inner ring nebula    around WR 16 suggests that the close surroundings of the WR star have been processed and enriched by molecular ejecta originated during previous LBV or RSG phases \citep{mwb99}, which very likely dominates the production of new CO   despite its depletion due the ionizing radiation field of the star. Then, component 1 may be  actually composed  by  a mix of stellar ejecta  and interstellar molecular gas.    This could mean that the X-factor used in Eq. \ref{eq:X} (1.6 $\times$ 10$^{20}$ cm$^{-2}$ (K \kms)$^{-1}$)   is high for this case, which originates some difference between $M({\rm H_2})_{\rm X}$ and  $M(\rm H_2)_{LTE}$. In fact, this might also explain the high ratio $T^*$$_{\rm R}$(CO)/$T^*$$_{\rm R}$($^{13}$CO)  observed for component 2 ($\sim$ 9;  see Table \ref{tabla:1}), which is almost twice the average Galactic value.     It is worth to point out that turbulent motions produce a slight drop in the value of X, although this effect is not significant \citep{bell06}.  Nevertheless, we keep in mind that $^{13}$CO LTE-based mass estimates are also subject to their own uncertainties and may underestimate true column density by a factor up to 7 \citep{pad00}. 

The study of higher rotational levels of the CO could shed  some light  on   the physical and excitation conditions of the molecular gas in the region.  Further, non-LTE analysis of these lines  will help to obtain  more robust estimates of some physical parameters, such as the column density, mass, kinetic temperature, and opacities.

\subsection{ Dynamics of the molecular gas     }

In previous sections we have detailed the characteristics of the molecular gas components projected onto the multiple ring nebula around WR 16. The good morphological correspondence between the CO  and  \ha\ emissions is  suggestive of  a  physical association between the molecular and the ionized gas.

 The morphology of the molecular gas in the velocity range from $\sim$ --10 \kms to --3 \kms suggests that  the ambient molecular gas  in the vicinity of  WR 16  could have been swept up by the action of  shock-fronts and the strong stellar winds, giving rise to the features shown in Fig. \ref{fig:comp1y2}.   Assuming that the winds of the star are spherically symmetric, and that component 1 and component 2  are two-dimensional projections of spherical expanding shells,  a connected structure with the velocity should be manifested in the datacube, first as a blueshifted pole, then into a growing and decreasing circular  ring, and finally to a redshifted pole, with the powering star placed at the center or close to it.  However,  according to our data none of these kinematical features are noticed, since no hints of a velocity gradient is detected.  As we have  pointed out, component 2 is very likely part of a larger molecular emission plateau that extends to the east (see Fig. \ref{fig:12y13}). We infer that WR 16 is placed outside of, or in an eccentric position with respect to this plateau, and is blowing the  molecular gas from a side (face-on).  Evidences of this kind of molecular structures have been found in the Galaxy (see \citealt{arce12}).   The location of the brightest region of the outer ring nebula, almost in the opposite direction of component 1 and component 2, suggests that this ring might be the remains of the parental molecular gas after been ionized by the WR star and/or its progenitor.  The molecular clump located at \radec\ $\approx$ \hbox{(9$^h$54$^s$,--57\gra 33\arcmin}) could be a remnant of the parental molecular gas (see Fig. \ref{fig:comp2halfa}).

%It  is also  worth mentioning that a molecular structure with similar morphology to component 2 was found to be associated with the optical ring nebula RCW 78  around the WN7 star HD 117688 (WR 55)\ \citep{crmr09,d&a12}. Evidences of this kind of molecular structures have been found in the Galaxy \citep{arce12}. 

As mentioned in Sect. 3.1, the   stratified location and kinematics of component 1 and component 2 (and probably component 3) is highly correlated with the  multiple layer  morphology of the nebula, and may be indicative of an evolution of the ISM associated with  successive shock-fronts, each one related with different evolutive episodes  of the predecessors of WR 16.    The different velocities of each component  could have been produced by violent events related to the star evolution and the probable occurrence of mass eruptions, which combined with the strong stellar UV radiation field, have produced the \ha\ morphology and the observed kinematical features.     Component 2 seems to be the more extended.    Its  border is adjacent to the eastern  border of the outer H$\alpha$ ring. Probably this is the oldest one and could represent the main-sequence signature of the star progenitor  (see below). Component 1 is  placed almost at the interface between component 2 and the H$\alpha$ emission. It  could be related to an intermediate phase prior to the actual WR stage, such as RSG or LBV \citep{mwb99}. Finally, the location of component 3 is internal to the other two components, and its  origin could be possible associated with a youngest event.    

The mentioned scenario is remarkably similar to that depicted by \citet{rmm01b,r203} for NGC 2359. In that case, however, the main ``fingerprint'' of the previous O-stage was revealed by the presence of a H{\sc i} shell expanding at 12 \kms that shocked the molecular gas accelerating it to a velocity of $\sim$ 12-13  \kms. In order to test the proposed scenario a rough estimate of  the dynamical age ($t_{\rm dyn}$) of component 2 was made using 
\begin{equation}
t_{\rm dyn}\ =\ 0.5 \times 10^6\ \ R / V_{\rm exp} \qquad \qquad  \textrm{(yr),}
\end{equation}
\citep{lc99}. Here,  $R$ is the averaged radius of component 2 ($\sim$ 9 pc at 2.3 kpc), and $V_{\rm exp}$ is the expansion  velocity. For $V_{\rm exp}$ we consider a conservative range of values from $\sim$5 \kms \citep{d&a12} to $\sim$12 \kms \citep{r203}. Then, the obtained values of $t_{\rm dyn}$   are in the range  \hbox{4 - 9 $\times$ 10$^5$} yr. If we consider the 35 M$\odot$ model of stellar evolution of \citet{gs96b} (O$\rightarrow$RSG$\rightarrow$WR), the duration of the O phase (4.5 $\times$ 10$^6$ yr) is larger  than  $t_{\rm dyn}$   by a factor of $\sim$ 5 - 10. On the other hand, taking into account the 60   M$\odot$ model of  \citet{la94} (O$\rightarrow$pre-LBV$\rightarrow$LBV$\rightarrow$WR) the duration of the O phase is larger only by  a factor $\sim$ 2 - 4.     In any case, these differences could be explained  in terms of the  small solid angle subtended by component 2 around WR 16 (see Fig. \ref{fig:comp2halfa}), which is not   able to  be impelled by the totality of the   stellar wind power.  This scenario allows one to deal  with the  hypothesis by which the parental molecular cloud around an evolved massive star is  destroyed and/or evacuated after  star formation. It is worth to point out that we have searched for H{\sc i} shells around WR 16, using available surveys, having negative results. This reinforces  the proposed scenario of component 2 being blown-up by the O-star progenitor.

The morphology and  location of component 1  might lead one to assume that this feature is composed  of  molecular gas ejected by the star in a previous evolutionary phase. However, the molecular mass estimated in Sect. 3.1  led us to conclude that this component  might be   rather  composed by  a mix of shocked interstellar  molecular gas  and stellar ejecta from intermediate stages (RSG or LBV), probably as a result of the encounter of two winds (as suggested by \citealt{gs96a,gs96b})   impinging onto the interstellar molecular gas. The most likely scenario is the WR wind encountering an already stopped RSG/LBV wind in contact with the molecular cloud, and being abruptly decelerated giving rise to the observed shocked gas. The stellar  ejecta  has been probably accumulating at the inner edge of component 1, as proposed by \citet{mwb99},    where the WR wind is impacting and   the UV field is ionizing,  giving rise to the inner \ha ring. %    Considering the velocity of the stellar wind during the WR stage ($\sim$ 2000 \kms) .......         
The velocity difference among component 1 and component 2 (2 - 3 \kms)  would give a rough estimate of the velocity of the shock, although this is very likely a lower limit since we are observing projected radial velocities from a face-on geometry. The location of component 1 entails that the wind of the previous O-stage have not fully interacted with the ISM and  could have partially run through the molecular gas. Indeed, this might be possible owing to the relatively low density of the O-stage winds. This argument was also discussed by  \citet{r203},  and this  scenario implies that only a fraction of the mechanical energy injected by the winds of the previous O-stage is converted into kinetic energy of the expanding molecular gas, which is in line with theoretical predictions   \citep{k92,ar07}.

\subsection{The inner ring}

In this section we will briefly discuss some aspects of the inner region of the nebula. In their Fig. 8, \citet{mwb99}  show the continuum 60 $\mu$m emission distribution around the star WR 16. The IRAS map  reveals a bright shell that has an excellent spatial correlation with the inner ring nebula  (no other external IR features were reported in the paper).     This shell is also detected in the HIRES IRAS 25 $\mu$m emission (see Fig. 9 of  \citealt{mwb99}).     We have analyzed the dust continuum emission at 60 $\mu$m  and 100 $\mu$m  from IRAS high resolution images (HIRES) to study the warm dust counterparts of the inner ring nebula. Following \citet{d90} the dust temperature of the shell was obtained using
\begin{equation}\label{td} 
\qquad T_{\rm d}\ =\ 95.94/\textrm{ln} \left[ 1.667^{(3+n)} (F_{100}/F_{60})  \right]   \qquad ({\rm K})
\end{equation}
 where  $F_{100}$ and $F_{60}$ are the 100 $\mu$m and 60 $\mu$m IRAS fluxes. The parameter $n$ (assumed to be 1.5)   is related to the absorption efficiency of the dust. We obtained a dust temperature of about T$_{\rm d}$ $\approx$ 28 K, which is remarkably similar to that found in the ring nebula NGC 2359 \citep{rmh01a}.   We also derived  the dust mass  of the shell using
\begin{equation}\label{}
M_d = 4.81\ \times\ 10^{-12} \ F_{100}\ d^2 \ (e^{143.88/T_{\rm d}} - 1) \  \ \ \ \ ({\rm M}_{\odot})
%M_d = m_n\  F_{60}\ d^2 \ (B_n^{2.5}\  -\ 1)                 \qquad ({\rm M}_{\odot}) 
\end{equation}
\citep{tok00}, where {\it d} the distance in pc, and $F_{100}$ is  given Jy. We obtain a dust mass of \hbox{$\sim$ 0.1} \msun with an uncertainty of the order of $\sim$ 45 $\%$. By assuming a standard gas-to-dust ratio of \hbox{$\sim$ 100}, and taking into account an ionized gas of $\sim$ 4 \msun (see Sect. 3.2), a molecular gas component of $\sim$ 6 \msun\ in mass is predicted to be associated with the dust shell around the WR star, which could explain the presence of component 3.

In Fig. \ref{fig:comp3}  we show an overlay of the mean CO line emission in a velocity range -9.59 \kms to -9.48 \kms (component 3)  onto the HIRES 60 $\mu$m emission which has been smoothed down to a common beam of  2.$'$7. A look at this figure shows that  the 60 \micr\ and molecular emission are asymmetrical, having most of their emission toward the north western section. 
\begin{figure}[h!]
\centering
\includegraphics[width=220pt]{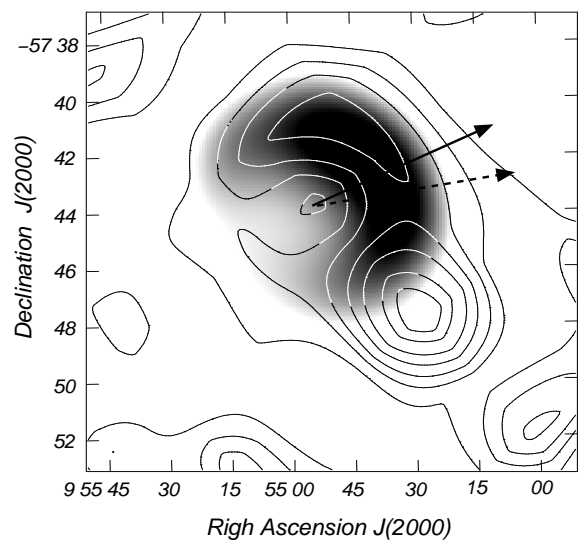}
\caption{  Overlay  of the mean CO line (contours) in the velocity range from --9.59 \kms to --9.48 \kms, and HIRES 60 $\mu$m emission (grey scale) smoothed down to a beam of 2$'$.7.  Dashed and solid arrows indicate the direction of the proper motion derived by \citet{mof98,mof99} and in this work, respectively.   }
\label{fig:comp3}
\end{figure}   
The same behavior is observed for the optical  and the radio continuum  emissions (see   Figs.  \ref{fig:wr16halfa}  and  \ref{fig:5ghz}). Using proper motions obtained from the Hipparcos astrometry satellite \citep{hip97}  \citet{mof98,mof99} suggest that WR 16 is a candidate star to be a runaway. The direction of the peculiar tangential proper motion of WR 16 (according to \citealt{mof98,mof99}) is shown in Fig. \ref{fig:comp3}    by a dashed arrow. Following \citet{mof98,mof99} but  using the Tycho-2 Catalogue \citep{hog00}   we derived for WR 16  a new direction of the peculiar tangential velocity \hbox{($\mu_{l, pec}$, $\mu_{b, pec}$)} with respect to its local ISM. Adopting  observed proper motions of \hbox{$\mu_{\alpha}$cos($\delta$) = (--6.8 $\pm$ 1.4) mas yr$^{-1}$} and  \hbox{$\mu_{\delta}$ = (5 $\pm$ 1.4) mas yr$^{-1}$}  we  derived \hbox{($\mu_{l, pec}$, $\mu_{b, pec}$)} = \hbox{(--12.1  $\pm$ 1.1)  mas yr$^{-1}$}, \hbox{(--3.5 $\pm$ 2.6)  mas yr$^{-1}$}. The obtained direction of the peculiar proper motion computed in this way is depicted in Fig. \ref{fig:comp3}    by a solid arrow. Our results confirm those of \citet{mof99} and very likely suggest that the strengthening in the IR, radio-continuum, optical, and probably molecular emission may be caused  by  the motion of the WR star with respect to the ISM (``bow-shock'').

From the explained above, it becomes clear the need of better evolutionary models which may explain the existence of dust and molecular gas at such close distance to a WR star. However, as pointed out in Sect. 3.1, it is important to stress that the existence of component 3 is still to be confirmed by high-resolution  and sensitive observations.

\section{Concluding remarks}

We have performed a study of the molecular gas in the multiple ring nebula associated with  the WN8 star WR 16.  In this work,  NANTEN \coa\ and  \cob\ observations   were used, together with 4.85 GHz radio continuum data retrieved from the PMNRAO Southern Radio Survey, and IRAS (HIRES) archival data.

%We  used NANTEN $^{12}$CO (J=1-0) observations (HPBW = 2'.7) of  region  $\sim$ 0\fdg58 in size around the star  obtained with a sampling of $\sim$2'.7, and  a second datacube $\sim$1\fdg6 in size  with a sampling of 5'.4 to analyze the molecular large-scale distribution. To study the ionized gas in the nebula,  we used radio-continuum data at 4.85 GHz obtained from the PMNRAO Southern Radio Survey. To account for the properties of the dust in the ring nebula we used archival IRAS (HIRES) images.

 The new CO and $^{13}$CO data allowed  us to map the whole molecular gas surrounding the star WR 16. We reported two molecular features at velocities of $-5$ \kms (component 1)   and $-$8.5 \kms (component 2). For component 1 we derive a molecular mass $\gtrsim$ 900 M$_{\odot}$ which indicates that is mostly composed by interstellar gas. We also   find that $M_{\rm vir}$ $>$  $M({\rm H_2})_{\rm X}$ $>$ $M(\rm H_2)_{LTE}$ which suggests that this feature is probably  binded by the external pressure of the ionized gas, and  is enriched by stellar molecular ejecta.  The morphological and kinematical stratification of the molecular gas in the velocity range from $\sim$ $-$10 \kms to $-$3 \kms is indicative of an evolution of the surrounding molecular gas with successive events associated with the evolution of the WR star.   Based on our findings, it is suggested that WR 16 could have been located close to the edge, or outside of a large molecular structure. The winds of the progenitor O-star could be responsible of the formation of component 2, ploughing part of the molecular gas surrounding the star. The rest of the  molecular gas was first  impacted by the winds of  previous RSG or LBV phase, and later by the WR winds, which originated a shock front, and the kinematical features observed in the CO emission.

The 4.85 GHz  emission shows a relatively good morphological correspondence with the optical nebula. The derived electron densities are about 16 \cm3 for  the inner nebula, and 8-10 \cm3 for  the outer nebula. The total ionized gas is $\sim$36 \msun. 

The dust shell around WR 16 reported by  \citet{mwb99} was studied using HIRES IRAS 60 $\mu$m and 100 $\mu$m. We determined a dust temperature and mass of about 28 K and 0.1 \msun, respectively. We also reported a third molecular feature at $-$9.5 \kms (component 3),  having a good morphological correlation with the the dust continuum emission at 60 $\mu$m and with the \ha emission, which has to be confirmed with high resolution observations. The asymmetry of the discovered structures may also be qualitatively explained by the motion of the massive star.

 \begin{acknowledgements}

 We especially thank  Prof. Takahiro  Hayakawa,  Prof. Yasuo Fukui, and Prof. Takeshi Okuda  for making  their   \cob\ data  available to us.  We very much acknowledge the anonymous referee for her/his helpful comments and suggestions which lead to the improvement of this paper.  We greatly appreciate the hospitality of all staff members of Las Campanas Observatory of the Carnegie Institute of Washington, and  all members of the NANTEN staff, in particular Prof. Yasuo Fukui, Dr. Toshikazu Onishi, Dr. Akira Mizuno, and students Y. Moriguchi, H. Saito, and S Sakamoto. We also would like to thank Dr. D. Miniti (Pont\'{\i}fica Universidad Cat\'olica, Chile) and Mr. F Bareilles (IAR) for their involvement in early stages of this project.

This project was partially financed by the Consejo Nacional de Investigaciones Cient\'ificas y T\'ecnicas (CONICET) of Argentina under projects  \hbox{PIP 112-200801-01299}, Universidad Nacional de La Plata (UNLP) under project \hbox{11G/091},  Agencia Nacional de Promoci\'on Cient\'ica y Tecnol\'ogica (ANPCYT) under project \hbox{PICT 14018/03}, and CONICYT Proyect PFB06.

\end{acknowledgements}

\bibliographystyle{aa}
\bibliography{bibliografia-wr16}
 
\IfFileExists{\jobname.bbl}{}
{\typeout{}
\typeout{****************************************************}
\typeout{****************************************************}
\typeout{** Please run "bibtex \jobname" to optain}
\typeout{** the bibliography and then re-run LaTeX}
\typeout{** twice to fix the references!}
\typeout{****************************************************}
\typeout{****************************************************}
\typeout{}

}

\end{document}